\documentclass{INTERSPEECH2023}

\DeclareMathOperator*{\argmin}{arg\,min}
\usepackage{multirow}
\usepackage{float}
\usepackage{placeins}

\interspeechcameraready

\title{Distillation Strategies for Discriminative Speech Recognition Rescoring}
\name{Prashanth Gurunath Shivakumar, Jari Kolehmainen, Yile Gu, Ankur Gandhe, Ariya Rastrow and Ivan Bulyko}
\address{
  Amazon Alexa AI, USA}
\email{\{psshvak,jkolehm,yilegu,aggandhe,arastrow,ibbulyko\}@amazon.com}

\begin{document}

\maketitle
 
\begin{abstract}
%Second-pass rescoring is employed in most state-of-the-art speech recognition systems.
%Recently, BERT based models have gained popularity for re-ranking the n-best hypothesis by exploiting the knowledge gained from masked language model pre-training.
%Further, fine-tuning with discriminative loss such as minimum word error rate (MWER) has shown to perform better than likelihood-based loss.
%Streaming applications with low latency requirements impose significant constraints on the size of the models, and thereby limiting the word error rate (WER) performance gains.
%In this paper, we propose effective strategies for distilling from large models discriminatively trained with the MWER objective.
%We train and evaluate our models on Librispeech and production scale internal dataset for voice-assistant.
%Our experiments demonstrate that our distillation techniques can provide relative improvements of upto 5.8\% in terms of WER over student models trained with MWER objective.
%We also show that the proposed distillation can reduce the WER gap between the student and the teacher by upto 93\%. 

Second-pass rescoring is employed in most state-of-the-art speech recognition systems.
Recently, BERT based models have gained popularity for re-ranking the n-best hypothesis by exploiting the knowledge from masked language model pre-training.
Further, fine-tuning with discriminative loss such as minimum word error rate (MWER) has shown to perform better than likelihood-based loss.
Streaming applications with low latency requirements impose significant constraints on the size of the models, thereby limiting the word error rate (WER) performance gains.
In this paper, we propose effective strategies for distilling from large models discriminatively trained with the MWER objective. 
We experiment on Librispeech and production scale internal dataset for voice-assistant. Our results demonstrate relative improvements of upto 7\% WER over student models trained with MWER. 
We also show that the proposed distillation can reduce the WER gap between the student and the teacher by 62\% upto 100\%.

\end{abstract}

\section{Introduction}
Two-pass automatic speech recognition (ASR) systems comprise of a first-pass model to generate n-best hypotheses and a second-pass rescoring model to re-rank and pick the best hypothesis.
Directly optimizing the second-pass models with respect to the WER is an appealing proposition.
One way to achieve this is through the MWER objective, which is designed to directly minimize the expected WER \cite{mwer2016,xu2022rescorebert}.
There have been several attempts at incorporating discriminative training using MWER loss for LSTM-RNN \cite{mwer2016,gandhe2020audio}, Sequence-to-Sequence \cite{mwer2018,li2020towards,sainath2019two}, RNN-Transducer \cite{guo2020efficient}, Transformer-Transducer \cite{meng2021minimum} based architectures for ASR.
Recent studies have shown BERT models \cite{bert} with bi-directional information encoding and pre-trained knowledge are favorable for rescoring, both in log-likelihood \cite{salazar2020masked} and discriminative loss \cite{xu2022rescorebert} settings.

Knowledge distillation in neural networks \cite{hinton2015distilling} has shown great success in efficiently compressing and mimicking one or ensemble of larger models (teachers).
This enables use of much smaller models with the advantages of reduced training time, inference and latency costs.
\cite{hinton2015distilling} proposed cross-entropy based distillation based on KL-divergence loss for distilling from an ensemble of teacher models for classification problems.
\cite{romero2014fitnets} extended knowledge distillation by minimizing L2-norm of the intermediate representations from hidden layers between the teacher and the student.
L2-norm based objectives are also used to closely replicate real value predictions of teacher in application to regression problems \cite{chen2017learning,wang2017model}.
There have also been several successful attempts of knowledge distillation in language modeling domain.
Most settings in the language model domain are posed as a classification problem and typically combination of the KL divergence loss with regularization is used.
For BERT models, optimizing the cosine distance between the teacher and student embeddings is found to be useful \cite{sanh2019distilbert}.
\cite{salazar2020masked} proposed to distill pseudo log-likelihoods (PLL) from BERT \cite{mlm-scoring1} using L2-loss based regression over the classification (CLS) token in application to utterance rescoring for ASR.

Recently, RescoreBERT was proposed~\cite{xu2022rescorebert} to train a BERT based second-pass rescoring model with several discriminative objective functions including MWER.
RescoreBERT provided up-to 5.3\% relative WER improvement, which was attributed to MWER fine-tuning.
The paper also proposed a technique to distill knowledge to a smaller model of 5M parameters for low-latency streaming applications.
The distillation involves training the student BERT model to predict the sentence-level PLL from the bigger 170M parameter teacher model before proceeding with MWER training.
Although, reducing the model size from 170M to 5M parameters brings average latency improvement of approximately 1350\% (relative), the WER improvements are diminished by 61.3\% relative to 170M model.
This suggests there is a need for better distillation strategies to attain better latency vs. WER trade-off.

The advantages of using discriminative loss functions such as MWER \cite{xu2022rescorebert,gandhe2020audio} to train language models are clear.
Further, the need for small models to enable low-latency streaming applications is necessary.
However, efficient techniques to distill from a model trained with discriminative loss such as MWER have not been explored.
%Moreover, the question on effect of distillation on the discriminative power of MWER trained LM is unanswered.
Moreover, the question on effect of distillation and its ability to retain the discriminative power with respect to MWER trained model is unanswered.
In this paper, we propose several techniques to perform distillation from a second-pass ASR rescoring teacher model trained with a discriminative MWER criterion.
We devise loss functions to train the student model to mimic the teacher while retaining the discriminative power of the teacher to minimize the expected WER.
We demonstrate that distilling from a larger teacher gives better WER in comparison to a MWER trained student model.
To the best of our knowledge, this work is the first to explore distillation from a MWER trained model and explore its feasibility in application to second-pass ASR rescoring.

%Although, in this paper, we apply the distillation techniques specifically to RescoreBERT, the proposed approach should generalize to any MWER trained LMs.

%The rest of the paper is organized as follows. We provide a brief description of RescoreBERT in section~\ref{sec:rescorebert}. Section~\ref{sec:distillation} presents the proposed distillation techniques. Section~\ref{sec:exp_setup} describes our experimental setup and datasets. Results are presented and discussed in section~\ref{sec:results}. Finally, the conclusion of the study is presented in section~\ref{sec:conclusion}.

\begin{figure}[t]
\begin{center}
    \includegraphics[width=0.75\columnwidth]{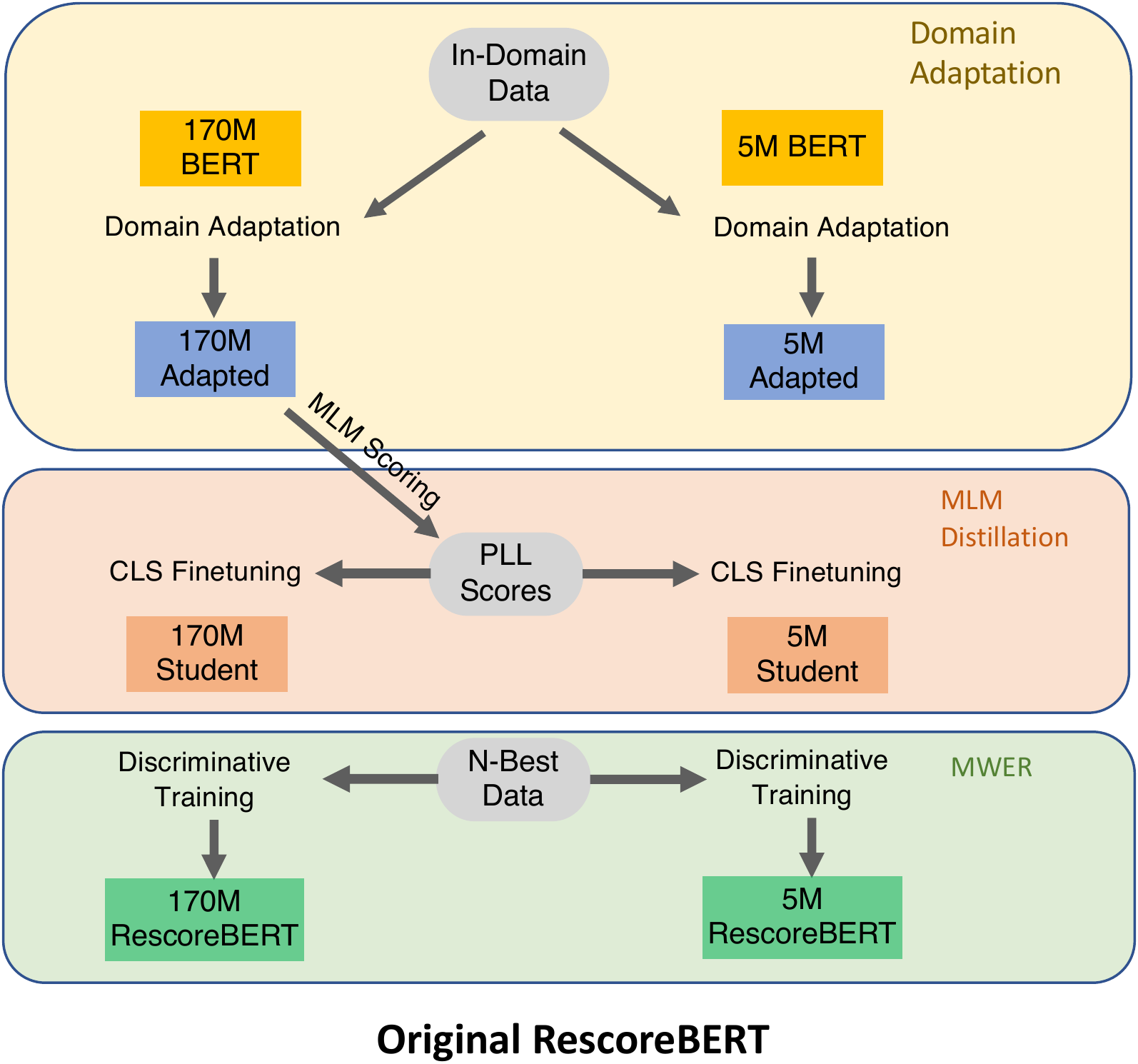}
    \caption{\centering {RescoreBERT Training Schema \cite{xu2022rescorebert} 
    \hspace{\textwidth}(boxes represent model training; rounded boxes represent data)}}
    \label{fig:rescorebert}
\end{center}
\vspace{-5mm}
\end{figure}

\section{RescoreBERT} \label{sec:rescorebert}
%This section provides a brief overview of RescoreBERT proposed originally in \cite{xu2022rescorebert}.
RescoreBERT is a discriminative second-pass rescorer based on BERT \cite{xu2022rescorebert}.
The architecture comprises of a BERT model with a feed-forward layer on the CLS token embedding.
The model consumes a sequence of tokens $X = (x_i,\ldots,x_N)$ and outputs a score.
Figure~\ref{fig:rescorebert} shows the schema for training RescoreBERT as proposed in \cite{xu2022rescorebert}.
The training process can be visioned in three stages: 
(i) Domain adaptation: pre-trained BERT model is further trained and adapted on the in-domain datasets using the masked language model (MLM) scheme \cite{bert}, (ii) MLM Distillation: this step involves addition of a feed-forward layer over CLS token embedding and fine-tuning to predict the PLLs \cite{salazar2020masked}:
\begin{equation}
\begin{split}
    PLL(X) &= -\sum_{t=1}^{N} \mathrm{log} P(x_t|X_{\backslash t}) \\
    & X_{\backslash t} = (\ldots,x_{t-1},\text{[MASK]},x_{t+1},\ldots)
    \end{split}
\end{equation}
and (iii) Discriminative training with MWER loss:
\begin{equation}\label{eq:mwer_loss}
    L_{MWER} = \sum_{j=1}^{n}(E_{j}-\overline{E}) Softmax(s_{j})
\end{equation}
where $E_{j}$ is the edit distance of $j^{th}$ hypothesis in an utterance, $\overline{E}$ is the mean edit distance computed over all hypotheses of n-best belonging to the utterance, $s_{j}$ is the final score of hypothesis $j$ obtained after linear interpolation of the first pass scores with the RescoreBERT score, and the $Softmax(.)$ gives the posterior distribution of the hypothesis over the $n$-best for an utterance.

Authors in \cite{xu2022rescorebert} additionally propose a smaller size model of 5M parameters for low-latency streaming.
While most of the training scheme remains identical to the bigger model, the paper proposes to distill the PLLs from the bigger teacher model onto the 5M model during the MLM distillation stage.

\section{Proposed Distillation}\label{sec:distillation}
\begin{figure}[t]
    \centering
    \includegraphics[width=0.6\columnwidth]{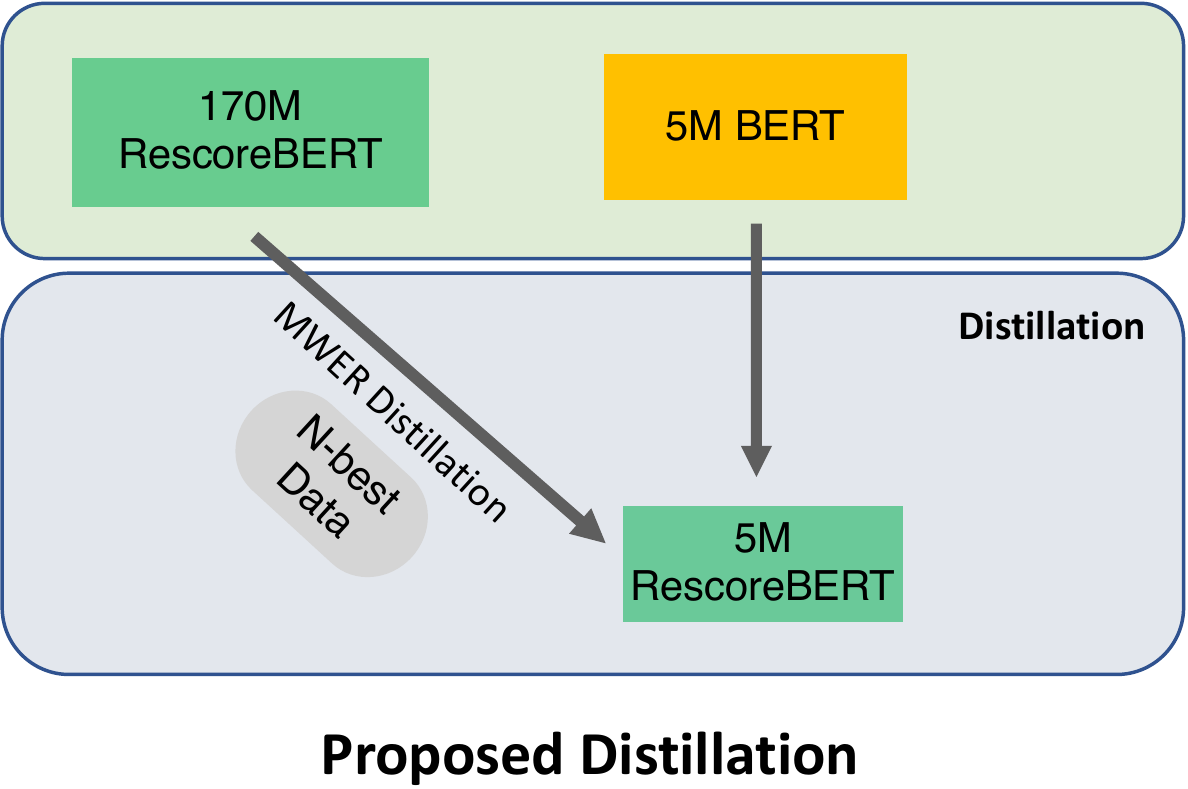}
    \caption{Proposed scheme for distilling from MWER models.}
    \label{fig:mwer_distillation}
\end{figure}

Study in \cite{xu2022rescorebert} noted that the low latency 5M model results in diminished WER gains relative to 170M teacher (upto 61.3\%).
Moreover, we hypothesize that the MLM distillation (where objective is language model scoring) is sub-optimal for the target task of rescoring. 
In this work, we propose to distill directly from the final MWER-trained teacher model as illustrated in Figure~\ref{fig:mwer_distillation}.
We present several loss functions for distillation to best preserve the discriminative power of MWER models.

\subsection{Cross-Entropy over n-best posterior distributions}
We propose to compute the cross-entropy between the posterior n-best distribution between the teacher and the student.
In order to compute the MWER loss objective (see Equation~\eqref{eq:mwer_loss}), we compute the softmax over the hypotheses over the n-best of the utterance.
We can directly optimize the student to mimic this distribution from the teacher, by using the following objective:
\begin{equation}\label{eq:ce}
L_{post} = CE(Softmax(s_{j}^s/T), Softmax(s_{j}^t/T))
\end{equation}
where $s_{j}^s$ is the predicted score from the student for hypothesis $j$ of an utterance, $s_{j}^t$ is the score from the teacher model for the same hypothesis, $T$ is the temperature parameter suggested in \cite{hinton2015distilling}, $CE(.)$ is the cross-entropy function and $Softmax(s_{j}/T) = \frac{exp(-s_{j}/T)}{\sum_{k=1}exp(-s_{k}/T)}$. 
%Note that this is equivalent to KL divergence between the student and teacher.

\subsection{Oracle based posterior correction}
One downside of the previous cross-entropy based distillation is that the upper-bound of the student is limited to that of the teacher.
\cite{hinton2015distilling} proposed having an additional cross-entropy loss against the groundtruth labels to account for instances when the teacher itself is sub-optimal.
However, since there is no ground-truth with PLL scores for MWER objective, we propose to use oracle hypothesis as the ground-truth \cite{variani2020neural} and add this as a correction term to the previously proposed loss in Equation~\eqref{eq:ce}:
\begin{equation}
    \label{eq:ce_oracle}
    \begin{split}
    L_{pOracle} &= L_{post} \\%\mathrm{CE}(Softmax(s_{j}^s/T), \mathrm{Softmax}(s_{j}^t/T)) \\
    &+ \alpha * CE(Softmax(s_{j}^s), OneHot(\argmin_j(E)))
    \end{split}
\end{equation}
where $OneHot(.)$ is the one-hot encoding function, and $\alpha$ is a tunable parameter.
The second term is estimating the oracle multinomial distribution over the n-best, whereas the first term mimics the distribution of the teacher.

\subsection{Mean Squared Error Loss}
We also explore the typical mean squared error (MSE) loss for distilling the score output directly from the teacher.
We experiment with two variations of the MSE loss for distillation.
First, where we sample hypothesis by maintaining the n-best structure (i.e., hypotheses belonging to an utterance appear in the same batch during training) and consider the 1st pass scores during optimization (we refer to this as $n$-best MSE):
\begin{equation}
    \label{eq:nbest_mse}
    L_{nMSE} = \frac{1}{n}\sum_{j=1}^{n}(s_{j}^{t}-s_{j}^{s})^2
\end{equation}
Second, by random sampling and including the scores from second-pass only (we refer to this as MSE):
\begin{equation}
    \label{eq:mse}
    L_{MSE} = \frac{1}{N}\sum_i^N(\hat{s}_i^{t}-\hat{s}_i^{s})^2
\end{equation}
$\hat{s}$ indicates the score is from second-pass only, $N$ is the batch size.
This method has the advantage of using additional text-only data during the distillation process.
\begin{table}[t]
    \centering
    \begin{tabular}{ll}
    \toprule
    Dataset & Hours/Tokens \\
    \midrule
    Domain adaptation training set & 3.8B tokens \\ %3,381,815,800 + 384038591 = 3,765,854,391 \\ %658M \\
    Domain adaptation development set  & 40M tokens \\% 25542366 + 14478950 = 40,021,316 \\ %7.5M \\
    N-best training set & 95K hours \\
    N-best development set  & 49 hours \\
    \midrule
    General & 194 hours \\
    Knowledge/Information Domain & 20 hours \\
    Navigation Domain & 39 hours \\
    Shopping Domain & 5 hours \\
    Music Domain & 40 hours \\
    Tail dataset & 100 hours \\
    \bottomrule
    \end{tabular}
    \caption{Data splits. Bottom half represent test-sets.}% $\dag$ indicates the test-set is synthetic.}
    \label{tab:data}
    \vspace{-5mm}
\end{table}

\section{Experiments}\label{sec:exp_setup}
\subsection{Data}
For experimentation, we use the publicly available Librispeech \cite{librispeech} and in-house dataset consisting of de-identified, far-field conversations with voice assistants for English.
Our in-house dataset is a compilation of multiple domains including Information/Knowledge, Navigation, Music, Video, Shopping and generic de-identified user interactions.
Partitions (non-overlapping) for training, development and testing are listed in Table~\ref{tab:data}.
%Two sets of training data is employed, one for domain adaptation and the other for MWER training.
Note, in the domain adaptation phase, we are not constrained to audio only data which allows us to use magnitudes more data for each of the domains.
Audio data is used for MWER training and distillation experiments.
%For MWER training, we are restricted to audio data for extracting n-best lists.
%The two training sets are not mutually exclusive.
%Finally, during the distillation phase, we use identical dataset as MWER training for fair comparison.
For testing purposes, we present the results on aforementioned domain specific datasets (see bottom half of Table~\ref{tab:data}).
The music domain dataset is a compilation of synthetic speech generated using text-to-speech (TTS) system.
The general test set is a compilation of de-identified user interactions with AI conversational agent and the tail dataset is a subset sampled from general test set and is curated based on tail linguistic data distribution.

\subsection{Experimental Setup}
\subsubsection{ASR system}
In case of Librispeech, we employ pre-trained, publicly available, tiny English Whisper model \cite{whisper} which is based on transformer sequence-to-sequence architecture, as the first pass.
For the internal dataset, the first pass model is based on RNN-T architecture \cite{he2019streaming} with shallow fusion n-gram language model \cite{gourav2021personalization,ravi2020improving}.
The RNN-T model is trained using the audio data listed in Table~\ref{tab:data} and also employs data augmentation using semi-supervised learning techniques.
The n-best size from the first pass is restricted to 10 for rescoring.

\subsubsection{Rescoring Model}\label{sec:exp_setup_RB}
Two BERT models differing in model sizes are used: (i) 170M parameter, and (ii) 5M parameter model.
The Teacher model comprises 16 layers with hidden size of 1024 and 16 attention heads (total of 170M parameters excluding embedding layer).
The student model comprises 4 layers with hidden size of 320 and 16 attention heads (total of 5M parameters excluding embedding layer).
Sentence-piece tokenizer trained on multilingual data \cite{raffel2020exploring} with a vocabulary size of 153,623 is adopted.

For Librispeech experiments, we use mini-BERT as student, base-BERT as teacher and closely follow the RescoreBERT setup in \cite{xu2022rescorebert}.
In case of internal dataset, the BERT pre-training is performed on MC4 dataset \cite{raffel2020exploring}.
Further, domain adaptation is performed on in-domain dataset as listed in Table~\ref{tab:data}.
Given larger data size, we found that the MLM distillation did not provide any improvements and hence skip it in this work.
Finally, a feed-forward layer is added to the CLS embedding and trained with MWER objective until convergence.
Interpolation weights for the first-pass is set to 20.0 and second-pass to 1.0 during training.
For run-time inference, the weights are optimized on the development dataset.

For the proposed distillation techniques, the 170M model described under Section~\ref{sec:exp_setup_RB} is used as the teacher to score each hypotheses in the n-best data.
The scores are distilled to the student whose architecture is identical to the baseline 5M model using losses proposed under Section~\ref{sec:distillation}.
%, however, we explore initializing from various stages, i.e., pre-domain-adaptation, pre-MWER-training and after MWER training.
%For pre-domain-adaptation and pre-MWER stages, the feed-forward layer is randomly initialized.
Parameter $\alpha$ in Equation~\eqref{eq:ce_oracle} is set to 1.0 for our experiments.

\subsection{Implementation}
%All the models are implemented in pytorch with DeepSpeed as the back-end and trained on 8 NVIDIA A100 GPUs.
We use a batch size of 48 during domain adaptation.
Batch size of 16 and 64 is used during MWER training for 170M and 5M model respectively.
Learning rate decay schedule is adopted with initial learning rate set to 1e-4 for 5M model and 1e-5 for the 170M model.
Adam optimizer is adopted during training.
The training is continued until convergence on the development dataset.
The baseline model, teacher and the distilled models are all trained on identical data for fair comparison.

\begin{table}[t]
    \centering
    \begin{tabular}{llll}
        \toprule
        \multirow{2}{*}{DataSets} & \multicolumn{3}{c}{\% WERR over 1st pass} \\
        \cmidrule{2-4}
                  & PLL 5M & Baseline 5M & Teacher 170M \\
        \midrule
        General & -1.15 & -3.16 & -4.20 \\
        Knowledge & -4.21 & -6.33 & -10.54 \\
        Navigation& -2.31 & -3.66 & -6.78 \\
        Shopping  & -1.82 & -7.03 & -10.82 \\
        Music     & -7.94 & -20.45 & -23.77 \\
        Tail      & -0.68 & -1.67 & -3.01 \\
        \bottomrule
    \end{tabular}
    \caption{WER relative improvements over first pass.}
    \label{tab:baseline}
    \vspace{-5mm}
\end{table}

%\begin{table*}[t!]
%    \centering
%    \begin{tabular}{lllllll}
%    \toprule
%    \multirow{2}{*}{Initialization stage} & \multicolumn{6}{c}{\% WERR on test-sets} \\
%    \cmidrule{2-7}
%     & Live test & Knowledge & Navigation & Shopping & Music$^{\dag}$ & Tail \\
%     \midrule
%    Before Domain Adaptation     & \textbf{-1.18} & \textbf{-2.36} & \textbf{-2.50} & -2.07 & -3.87 & -0.96 \\
%    After Domain Adaptation      & -0.79 & -2.02 & -2.03 & \textbf{-3.59} & \textbf{-5.49} &  -0.96 \\
%    After MWER Training (Baseline) & -0.99 & -2.02 & -1.59 & -3.37 & -4.72 & -0.83 \\
%    \bottomrule
%    \end{tabular}
%    \caption{Results of rescoring models in terms of word error rate reduction (WERR) relative to the baseline 5M model on different test-sets. Negative numbers indicate WER improvements and positive numbers indicate degradation. $\dag$ indicates the test-set is synthetic.}
%    \label{tab:init}
%    \vspace{-5mm}
%\end{table*}

\begin{table*}[t]
    \centering
    \begin{tabular}{lllllll}
    \toprule
    \multirow{2}{*}{Model} & \multicolumn{6}{c}{\% WERR on test-sets} \\
    \cmidrule{2-7}
     & General & Knowledge & Navigation & Shopping & Music$^{\dag}$ & Tail \\
     \midrule
%    Teacher 170M     & -1.38 & -4.71 & -3.90 & -3.70 & -7.58 & -1.24 \\
%    MSE & -0.79 (57.2) & -2.02 (42.9) & -2.03 (52.1) & -3.59 (97.0) & -5.49 (72.4) & 0.96 (77.4) \\
%    n-best MSE & -1.1 (79.7) & -2.52 (53.5) & -1.05 (26.9) & -2.66 (71.9) & -5.26 (69.4) & -0.76 (61.3) \\
    Teacher 170M            & -1.19 & -4.5 & -3.38 & -4.55 & -7.72 & -1.39 \\
    $L_{MSE}$                     & -0.40 (33.6) & -1.93 (42.9) & -1.27 (37.6) & -3.79 (83.3) & -5.79 (75.0) & -1.25 (89.9) \\
%    n-best MSE              & -0.79 & -1.61 & -0.84 & -2.49 & -5.56 & -0.7 \\
    $L_{nMSE}$              & -0.79 (66.4) & -2.25 (50.0) & -1.27 (37.6) & -3.90 (85.7) & -5.56 (72.0) & -1.11 (79.9) \\
%    CE (T=1)                & -0.79 (66.4) & -2.57 (57.1) & -0.70 (20.7) & -1.41 (37.2) & -5.48 (71.0) & -0.84 (75.7) \\
    $L_{post}$                & -0.99 (83.2) & -3.22 (71.6) & \textbf{-2.11 (62.4)} & -2.60 (57.1) & -5.56 (72.0) & -1.11 (79.9) \\
    $L_{pOracle}$           & -0.99 (83.2) & -2.25 (50.0) & -1.13 (33.4) & -3.25 (71.4) & -4.78 (61.9) & -1.11 (79.9) \\
    %$L_{pOracle}$ (T=2)           & -0.4 & -2.24 (50.0) & -0.14 & -1.95 & -5.25 & -0.7\\
%    CE (T=3)                & +0.79 & -0.32 & -0.14 & 0.0 & -6.48 & +0.14 \\
%    CE + Oracle correction  & & 
    %$L_{nMSE} + L_{MWER}$       & -0.79 (66.4) & -2.89 (64.2) & -1.41 (41.7) & -3.79 (83.3) & -5.71 (74.0) & \textbf{-1.39 (100.0)} \\
    $\beta L_{nMSE} + L_{MWER}$   & -0.79 (66.4) & -2.57 (57.1) & -1.83 (54.1) & \textbf{-4.22 (92.7)} & -5.63 (72.9) & -1.25 (89.9) \\
    %$L_{post}$ (T=1) + $L_{nMSE}$   & \textbf{-1.19 (100.)} & -2.89 (64.2) & \textbf{-2.39 (70.7)} & -3.79 (83.3) & -5.25 (68.0) & -1.11 (79.9) \\
    $L_{post}$ + $\beta L_{nMSE}$ & \textbf{-1.19 (100.)} & -2.24 (50.0) & -1.13 (33.4) &  -3.03 (66.6) & -5.17 (67.0) & \textbf{-1.39 (100.)} \\
    $L_{post}$ + $\beta L_{MWER}$ & -0.79 (66.4) & \textbf{-3.85 (85.5)} & -1.27 (37.6) & -3.90 (85.7) & -5.09 (65.9) & -1.25 (89.9) \\
    $\gamma L_{post}+\beta L_{nMSE}+\gamma L_{MWER}$ & -0.79 (66.4) & -3.21 (71.5) &  -1.55 (45.9) & -3.57 (78.5) & \textbf{-7.02 (90.9)} & -1.25 (89.9) \\
    \bottomrule
    \end{tabular}
    \caption{Results of rescoring models in terms of word error rate reduction (WERR) relative to the baseline 5M model on different test-sets. \dag indicates the test-set is synthetic. Numbers inside parenthesis are percentage improvements from distillation relative to WER gap between teacher and baseline. $\beta$=0.01;$\gamma$=0.5. All the improvements are statistically significant with $p<0.002$ (Matched Pair Sentence Segment - Word Error \cite{gillick1989some}) over the baseline.}
    \label{tab:results}
    \vspace{-5mm}
\end{table*}

\subsection{Baseline and Evaluation}
We use two baseline systems in this study: (i) the 5M parameter RescoreBERT model proposed in \cite{xu2022rescorebert} (see Section~\ref{sec:exp_setup_RB}), and (ii) the PLL based rescoring model distilled from 170M domain adapted MLM \cite{salazar2020masked}.
The PLL baseline is used to verify claims that the proposed distillation technique is able to preserve the discriminative power of MWER trained teacher.
The 5M RescoreBERT baseline is used for all other experiments.
%The first pass model is maintained as a constant throughout all the experiments.

Evaluations are performed on unseen held-out test sets listed in Table~\ref{tab:data}.
We present results in terms of percentage word error rate reductions (WERR) over the baseline 5M model.
We would like to note that the absolute WER is less than 13\% for synthetic datasets and well below 10\% for others.
In order to assess the effectiveness of each of the distillation techniques, we also provide the percentage improvement relative to WERR from teacher over 5M baseline model.

\section{Results}\label{sec:results}
We first perform experiments on the internal datasets and then use the best distillation technique to present results on Librispeech.
Table~\ref{tab:baseline} shows the performances of the baseline systems and the 170M RescoreBERT teacher model on internal datasets.
Compared to using PLL, the benefits of MWER training is clear.
There is substantial performance gap between the baseline 5M and the teacher attributed to bigger model size.

\subsection{Initialization strategy for distillation}
Next, we performed experiments using vanilla MSE loss ($L_{MSE}$) to deduce the optimal initialization point for the student model before distillation.
We consider, three candidate initialization points: (i) before domain adaptation, i.e., distilling directly to BERT model, (ii) after domain adaptation and before MWER training, and (iii) after MWER training, i.e., using the baseline 5M RescoreBERT as starting point for distillation.
%Table~\ref{tab:init} lists the results in terms of percentage WERR relative to 5M RescoreBERT baseline.
%Firstly, we find that distillation improves WER over the baseline model \cite{xu2022rescorebert} across all the test-sets.
%Secondly, we observe initializing on MWER trained baseline provides no benefits, i.e., performing MWER training before distillation is not beneficial.
%Third, we find distillation without domain adaptation results in a large performance gap for synthetic datasets (see Music in Table~\ref{tab:init}).
We observed initializing on MWER trained baseline provides no benefits, i.e., performing MWER training before distillation is not beneficial.
%Moreover, distillation without domain adaptation results in a large performance gap for synthetic datasets (Music in Table~\ref{tab:init}).
Moreover, distillation without domain adaptation results in a large performance gap for synthetic datasets (Music).
Considering the above, we find domain adapting the BERT model a good starting point for distillation and make this a standard for the rest of our experiments.

\begin{table}[b]
\vspace{-5mm}
    \centering
    \begin{tabular}{lll}
    \toprule
        Model &  test-clean & test-other \\
    \midrule
        Whisper \cite{whisper} & 5.67 & 12.89 \\
        Teacher (base-BERT) & 4.48 & 10.65 \\
        Baseline (mini-BERT) & 5.32 & 12.02 \\
        Distilled (mini-BERT) & \textbf{5.21 (13.1\%)} & \textbf{11.90 (8.8\%)} \\
    \bottomrule
    \end{tabular}
    \caption{WER Results on Librispeech. Numbers inside parenthesis are percentage improvements from distillation relative to WER gap between teacher and baseline.}
    \label{tab:librispeech}
    \vspace{-5mm}
\end{table}

\subsection{Comparison of distillation strategies}
Table~\ref{tab:results} presents the relative WER improvements obtained using proposed distillation techniques over the 5M baseline model.
We consider the 170M teacher as the upper limit and evaluate performance improvements obtained from distillation relative to reducing the gap between teacher and student.
Firstly, we find vanilla MSE loss can provide decent improvements over the MWER trained baseline.
The MSE loss is particularly effective on synthetic music dataset.
We believe this is because of the peculiar acoustic characteristics of the TTS generated signals, which can reduce reliance on the first pass scores.
Note, this is the only loss that doesn't account for the first pass scores explicitly.
%Next, we include the first pass scores for MSE and perform structured sampling preserving the n-best.
Inclusion of the first pass scores ($L_{nMSE}$) improves performance on datasets where the initial room for improvement is small (although we observe slightly smaller improvements on tail).
Both the MSE objectives are effective on datasets where the gap in student-teacher performance is large.
The cross-entropy ($L_{post}$) loss over posterior n-best distribution gives further improvements especially on the general, knowledge and navigation test-sets.
Optimum results were obtained with temperature set to 2.
We find the oracle correction ($L_{pOracle}$) provides benefits on shopping domain, however fails to improve on others.
%we obtained best results with temperature set to 2.
%The cross-entropy distillation scheme gives further improvements especially on the live, knowledge and navigation test-sets.
Additionally, we also experiment with linear combinations of MWER, MSE, cross-entropy loss and find it to be better than distilling with single loss objective.
%We find that the addition of MWER or the CE loss to MSE is strictly better than using only MSE on all the datasets (except for the synthetic test-set).
%We find that the best results are obtained with multi-loss objective.

Overall, proposed distillation helps reduce the WER gap between the 170M and 5M models by up-to 93\% relative on domain specific datasets.
We observe that the distillation is particularly effective for the general and tail data, recovering 100\% of teacher performance.
The performance of the teacher highlights the modeling challenge for tail data given minimal improvements.
This result shows that the distillation is not sensitive to the underlying data distribution.
%Results on domains such as Knowledge and Navigation show potential room for improvement with respect to the teacher.
%Finally, our best distilled model ($L_{post}$) provides improvements of 4.2\%, 9.03\%, 5.83\%, 10.11\%, 21.33\% and 2.74\% over first pass on General, Knowledge, Navigation, Shopping, Music and Tail test-sets respectively.
Finally, our best distilled model ($L_{post}$) provides  4.2\%, 9.03\%, 5.83\%, 10.11\%, 21.33\% and 2.74\% improvement over first pass on General, Knowledge, Navigation, Shopping, Music and Tail test-sets respectively.

\subsection{Results on Librispeech}
Table~\ref{tab:librispeech} lists results obtained using the best performing distillation scheme on Librispeech.
The teacher RescoreBERT gives a relative improvement of 21\%/17.4\% on test-clean/test-other over the first-pass.
The baseline RescoreBERT is able to achieve 6.2\%/6.7\% WERR on test-clean/test-other.
The distillation is able to reduce the teacher-baseline gap by 13.1\%/8.8\% on test-clean/test-other showcasing the effectiveness of the approach on different domains and different first pass models.

\section{Conclusions}\label{sec:conclusion}
In this study, we proposed new techniques to distill from a teacher trained with the MWER criterion. 
A cross-entropy based distillation loss computed over n-best posterior distribution, an oracle based correction to the cross-entropy loss and the traditional MSE loss were explored and compared to deduce the best approach for distillation.
We designed experiments and presented evaluations on publicly available Librispeech and domain specific internal datasets.
We showed that distilling from a larger teacher model, discriminatively trained on MWER loss, can outperform typical MWER training for the student.
Improvements of up-to 7\% on WER was obtained over the MWER trained student model.
The distillation is shown to reduce the WER gap between the teacher and student by 62\% upto 100\%.
%We also explore the best initialization strategies before distillation.
In the future, we plan to distill from a bigger teacher to maximize the impact of distillation.

%\section{Acknowledgements}

\bibliographystyle{IEEEtran}
\bibliography{mybib}

\end{document}